\documentclass[reprint,superscriptaddress,showpacs,bibnotes,amsmath,amssymb,aps,prl]{revtex4-1}

\usepackage{graphicx}
\usepackage{bm}

\begin{document}
\title{Charge frustration in a triangular triple quantum dot}

\author{M. Seo}
\affiliation{Department of Physics, Pusan National University, Busan 609-735, Republic of Korea}

\author{H. K. Choi}
\affiliation{Braun Center for Submicron Research, Department of Condensed Matter Physics, Weizmann Institute of Science, Rehovot 76100, Israel}

\author{S.-Y. Lee}
\affiliation{Department of Physics, Pusan National University, Busan 609-735, Republic of Korea}

\author{N. Kim}
\affiliation{Korea Research Institute of Standard and Science, Daejeon 306-600, Republic of Korea}

\author{Y. Chung}
\email[]{ycchung@pusan.ac.kr}
\affiliation{Department of Physics, Pusan National University, Busan 609-735, Republic of Korea}

\author{H.-S. Sim}
\affiliation{Department of Physics, Korea Advanced Institute of Science and Technology, Daejeon 305-701, Republic of Korea}

\author{V. Umansky}
\affiliation{Braun Center for Submicron Research, Department of Condensed Matter Physics, Weizmann Institute of Science, Rehovot 76100, Israel}

\author{D. Mahalu}
\affiliation{Braun Center for Submicron Research, Department of Condensed Matter Physics, Weizmann Institute of Science, Rehovot 76100, Israel}

\date{October, 12, 2012}

\begin{abstract}
We experimentally investigate the charge (isospin) frustration induced by a geometrical symmetry in a triangular triple quantum dot. We observe the ground-state charge configurations of six-fold degeneracy, the manifestation of the frustration. The frustration results in omnidirectional charge transport, and it is accompanied by nearby nontrivial triple degenerate states in the charge stability diagram. The findings agree with a capacitive interaction model. We also observe unusual transport by the frustration, which might be related to elastic cotunneling and the interference of trajectories through the dot. This work demonstrates a unique way of studying geometrical frustration in a controllable way.
\end{abstract}

\pacs{}
\maketitle
Highly degenerate ground states in a many-body system show interesting properties by symmetry and fluctuations. A related example is the geometric frustration of a triangular spin lattice \cite{ong2004electronic}. The phenomenon was first introduced by Linus Pauling to explain the residual entropy observed in water ice at absolute zero temperature \cite{pauling1935structure}. Later on, exotic many-body phenomena, induced by the geometric frustration, such as spin ice \cite{bramwell2001spin, ramirez1999superconductivity}, spin liquid \cite{balents2010spin,pratt2011magnetic} and spin ice magnetic monopole \cite{bramwell2009measurement} were observed. However these phenomena have been studied mainly in ensemble systems.

Quantum dots (QDs) provide an ideal platform for studying degenerate many-body ground states in a systematic way, as their parameters can be tuned in-situ \cite{meirav1990single,PhysRevLett.77.3613,livermore1996coulomb}. Degenerate ground states lead to Coulomb blockade resonances and Kondo effects in a single QD \cite{goldhaber1998kondo, van2000kondo}, and they are useful for manipulating qubits and quantum entanglement in a double QD \cite{koppens2006driven,PhysRevLett.107.146801,shulman2012demonstration}. The research has been recently extended to triple quantum dots (TQDs). There have been experimental reports on the TQDs of serial or asymmetric triangular geometry, which focus on charge rectification, Aharonov-Bohm effect, and coherent spin control \cite{vidan2004triple,schroer2007electrostatically,delgado2008spin,nowack2007coherent,PhysRevB.82.075304,gaudreau2011coherent}.
\begin{figure}[hb!]
\includegraphics[width=8.5cm]{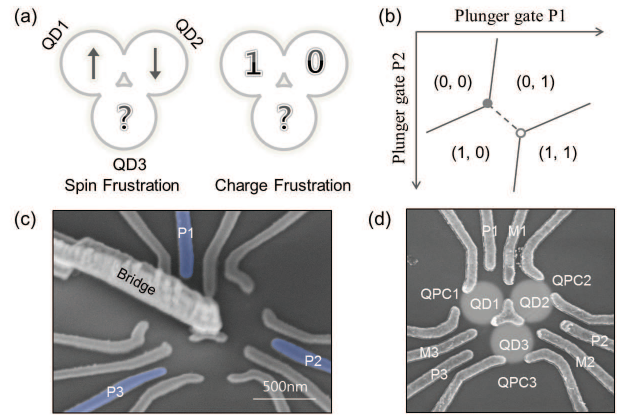}%
\caption{(a) Schematic view of geometric frustration in a TQD with antiferromagnetic coupling between spins, or Ising isospins by the charge degree of freedom. QD1 and QD2 are in up and down spin (or isospin) states, respectively. Then, due to the antiferromagnetic coupling and the geometric symmetry, the spin up (down) state of QD1 (QD2) forces QD3 to have spin down (up). Hence, the spin of QD3 cannot be determined and is frustrated (as denoted by the question mark). (b) Charge stability diagram of a double QD. On the dashed line connecting electron (filled circle) and hole (open circle) triple degeneracy points, there are two degenerate ground charge states of (1,0) and (0,1). These are described by antiferromagnetic isospin coupling. (c) and (d) SEM images of a symmetric TQD fabricated on a GaAs/AlGaAs 2DEG wafer. The 2DEG is buried 77 nm below the surface of the wafer. The carrier density is $1.9 \times 10^{11}$ cm$^{2}$, and the mobility is $1.1 \times 10^{6}$ cm$^{2}$/Vs at 4.2 K. The TQD is defined by 15/30 nm thick Ti/Au metallic gates, which is patterned by electron-beam lithography.
\label{fig1}}
\end{figure}

A symmetric triangular triple quantum dot is of interest since the geometric frustration can be realized in a single triangular lattice. Such realization will offer many advantages over ensemble systems, since the system can be precisely controlled experimentally and the intrinsic properties of the frustration, which might be hidden by ensemble average, can be found. 

In this work, we experimentally realize a symmetric TQD, and observe the ground-state charge configurations of $\textit six$-$\textit fold$ degeneracy, for the first time, by measuring electron transport; the six fold is the highest degeneracy realizable in a TQD. The degenerate ground states are the manifestation of $\textit charge$ $\textit frustration$, namely, the frustration of Ising isospins. We reveal the charge transport properties of the charge frustration. The six-fold degenerate states show omnidirectional transport among three reservoirs, each coupled to a dot of the TQD. They are accompanied by nearby nontrivial triple degenerate states in the charge stability diagram. These properties are understood, based on a capacitive interaction model. We also report unusual features of charge transport by the frustration, which might be partially related to elastic cotunneling and interference.

The frustration occurs when there is antiferromagnetic coupling between the dots of the TQD, as in Fig.~\ref{fig1}(a), when two dots have opposite spins to each other, the spin state of the other dot is frustrated. Even though it is highly interesting to realize such spin frustration states, experimental implementation is not trivial due to the difficulties of controlling electron spins in quantum dots.
\begin{figure}
\includegraphics[width=8.5cm]{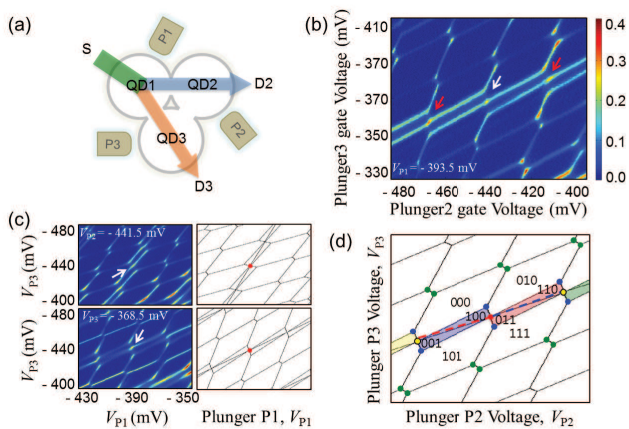}
\caption{(Color online) (a) The measurements are done by measuring current flow from QD1 to QD2 and QD3. (b) The conductance, measured with varying $V_{P2}$ and $V_{P3}$, constitutes the stability diagram on the P2-P3 plane. $V_{P1}$ is fixed at the value where a six-fold degeneracy point (marked by a white arrow) appears.  (c) The same, but on different planes of the plunge gates. (d) Stability diagram on the P2-P3 plane, calculated from Eq.~(\ref{eq:eq1}). It shows the charge frustration points (red and yellow circles), the triple degeneracy points of QD1 and QD2 (blue), and the triple points of QD2 and QD3 (green). The occupation numbers in the stability diagram are labeled such as (0,0,0), for clarity, by subtracting arbitrary constant numbers from the actual electron occupation numbers (which are positive) in TQD.\label{fig2}}
\end{figure}
Alternative way of studying geometric frustration is to use ($\textit n_{1}$,$\textit n_{2}$) = (1,0) and (0,1) degenerate charge states (dashed line in Fig.~\ref{fig1}(b)) of a double quantum dot, where $\textit n_{i}$ is the occupation number of QD $\textit i$. These states can be considered as two Ising isospins with antiferromagnetic coupling; for example, (1,0) is interpreted as isospin up in QD1 and down in QD2. By establishing the antiferromagnetic coupling between any two neighboring dots, the isospin frustration can be realized in a TQD and we call this situation as $\textit charge$ $\textit frustration$. In this situation, there occur six-fold degenerate ground states of ($\textit n_{1}$,$\textit n_{2}$,$\textit n_{3}$) = (1,0,0), (0,1,0), (0,0,1), (1,1,0), (1,0,1), (0,1,1). We remark that the six-fold degeneracy is the highest among the possible degeneracies in a TQD; here we do not count spin degeneracy. The advantage of using such charge states is that the isospins can be precisely controlled by plunger gate voltages. Note that the six-fold degeneracy has not been explicitly discussed even theoretically. 

The electrostatic energy $\textit E$ of a TQD can be described by a capacitive interaction model \cite{schroer2007electrostatically},
\begin{equation}
E(n_{1},n_{2},n_{3})=\sum_{i=1,2,3}U_{i}Q_{i}^{2}+\sum_{i\ne j}X_{ij}Q_{i}Q_{j}
\label{eq:eq1}
\end{equation}
Where $U_{i}$ is the intradot capacitance energy of QD $\textit i$, X$_{ij}$ is the interdot interaction between QDs $\textit i$ and $\textit j$, $Q_{i} = n_{i} -\sum_{j} c_{ij}V_{j}$ is the excess charge in QD $\textit i$, $V_{i}$'s are plunger gate voltages, and $c_{ij}$'s are coupling coefficients. Single-particle level spacing and Zeeman energy are ignored. The six-fold degeneracy appears when the interdot interactions have the same strength, $X_{ij}=X$.

Figure~\ref{fig1}(c) shows a symmetric triangular TQD fabricated on a GaAs/AlGaAs 2DEG wafer. Each dot of the TQD couples with a reservoir in the tunneling regime. The dot-reservoir tunneling is controlled by three QPC gates (QPC-$\textit i$, $\textit i$=1,2,3), and the interdot tunneling is controlled by coupling gates (M-$\textit i$) and a center gate with a bridge structure. The six-fold degeneracy condition of $X_{ij}=X$ is achieved, by $\textit iteratively$ tuning the QPC gates and the coupling gates. Since this iteration process requires to measure many 3D stability diagrams, we used a homemade wide-band low-noise current amplifier, which is capable of taking 20 conductance data points per second \cite{kretinin2012wide}.

The six-fold degeneracy (charge frustration) point is confirmed by measuring zero-bias electron differential conductance. Figure~\ref{fig2} shows the charge stability diagrams, obtained by measuring the total current from QD1 to the other two dots of QD2 and QD3; the current from QD2 or QD3 shows qualitatively the same results. The measured diagram agrees with the computation based on Eq.~(\ref{eq:eq1}); see Figs.~\ref{fig2}(b) and~\ref{fig2}(d). The comparison shows $U_{i} \sim$ 0.27 meV and $\textit X \sim$ 0.06 meV in our TQD. We note that spin states are not resolved at our base temperature.

Figures~\ref{fig2}(b) and~\ref{fig2}(d) show the measured and calculated stability diagrams in the P2-P3 plane of the three dimensional P1-P2-P3 diagram. The charge configurations around the red point in Fig.~\ref{fig2}(d) confirm that the six different ground states of (1,0,0), (0,1,0), (0,0,1), (1,1,0), (1,0,1), (0,1,1) are indeed degenerated on the point. In the plane, a series of such six-fold points (yellow dots) appear periodically, implying that the TQD is highly triangular symmetric. The six-fold points are also observed in other planes (P1-P2, P1-P3) as shown in Fig.~\ref{fig2}(c). The mismatch in three voltage coordinates ($V_{P1}$, $V_{P2}$, $V_{P3}$) of frustrated points (three red points in Figs.~\ref{fig2}(c), (d)) is less than 0.7 mV ( $\sim$ 8.2 $\mu$eV in energy), which is comparable to 2$k_{B}$T ( $\sim$ 9 $\mu$eV) at 52 mK of our base temperature.

\begin{figure}
\includegraphics[width=8.5cm]{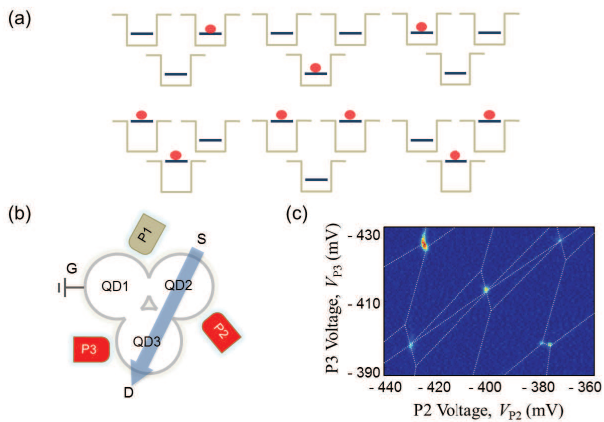}%
\caption{(Color online) (a) Energy diagrams of the six-fold degenerate ground-state charge configurations on the charge frustration point of a TQD. A red circle represents an electron occupying a dot, and a blue line shows the electrochemical potential of a dot. The six different configurations have the same electrostatic energy, resulting in omnidirectional transport via the degenerate states. (b) Measurement setup and (c) data of conductance through QD2 and QD3. The boundary lines (dotted lines) of the stability diagram are shown as guidelines.\label{fig3}}
\end{figure}

On the six-fold points, the charge frustration implies the maximal charge fluctuations without energy cost, hence shows good conductance. Figure~\ref{fig3}(a) shows the energy diagrams of a TQD on the charge frustration point. When only one electron occupies the TQD, the chemical potential of each dot of the TQD lies below the Fermi energy of the reservoirs. When two electrons occupy the TQD, the chemical potentials are aligned to the reservoir Fermi level. The resulting tunneling processes resemble the well-known sequential tunneling of a single quantum dot, and give rise to omnidirectional transport among the three reservoirs without energy cost, i.e., transport between any two of the three reservoirs. This is an important feature of the frustration, and it is confirmed by our measurement.
The observation of the omnidirectional transport is not the sufficient evidence for the charge frustration since the quadruple degeneracy point \cite{schroer2007electrostatically} in an asymmetric TQD shows a similar characteristic. Another distinct feature of the frustration is the nontrivial triple points (blue points in Fig.~\ref{fig2}(d)) located in the vicinity of the six-fold point. The existence of such triple points distinguishes the six-fold point from the quadruple point. These triple points are nontrivial in the sense that they are the triple points of QD1 and QD2 (rather than those of QD2 and QD3) in the P2-P3 plane where $V_{P2}$ and $V_{P3}$ vary, namely that the charges of QD1 and QD2 fluctuate on these points. Hence, electron transport through QD1 and QD2 on these points is expected, while the current flow through QD3 is Coulomb blockaded. Such directional current flow can be checked by measuring the conductance through QD2 and QD3 while grounding QD1 as shown in Fig.~\ref{fig3}(b). In this measurement scheme, the nontrivial triple points around the six-fold point will not contribute to the conductance. Experimental results in Fig.~\ref{fig3}(c) agree with this expectation. They show negligible current flow through QD3 on the nontrivial triple points in the P2-P3 plane, while meaningful flow on the six-fold points and on the trivial triple points (green points of Fig.~\ref{fig2}(d)) of QD2 and QD3. This confirms that the observed point is indeed a six-fold degeneracy point.

\begin{figure}
\includegraphics[width=8.5cm]{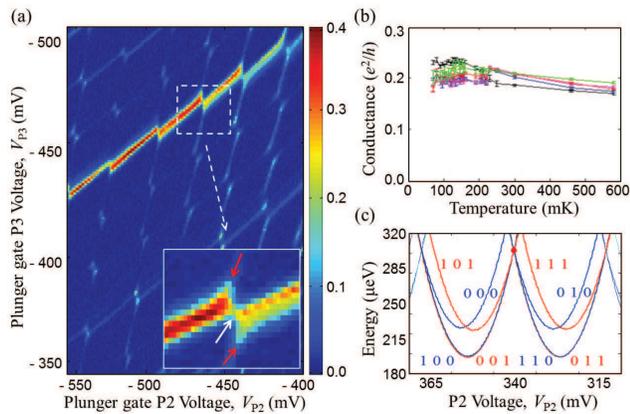}
\caption{(Color online) (a) Stability diagram, around six-fold degeneracy points, in a situation where electron tunneling between the TQD and the reservoirs is weaker than the case of Fig.~\ref{fig2}(b). The diagram is obtained by measuring conductance from QD1 to QD2 and QD3; see Fig.~\ref{fig2}(a). The highly conducting stripes are shown between two six-fold degeneracy points. Inset: The stability diagram inside the dashed box in the figure. The maximum value of the conductance in the stripe is higher than the values on the neighboring six-fold degeneracy point (white arrow) and triple points (red arrows). (b) Temperature dependence of the conductance at various points in the stripes. (c) The calculated electrostatic energies of the TQD along the border between (0,0,1) and (1,0,0) [dashed red line in Fig.~\ref{fig2}(d)].\label{fig4}}
\end{figure}

Next, we turn back to the conductance from QD1 to QD2 and QD3 [see Fig.~\ref{fig2}(a)] in the P2-P3 plane, and focus on another nontrivial feature of the charge frustration in the domains (0,0,1) and (1,0,0) around the six-fold points [see the shaded boxes in Fig.~\ref{fig2}(d)]. In Fig.~\ref{fig2}(b), these regions exhibit much weaker conductance signals than the six-fold points, as expected. However, when electron dot-reservoir tunneling becomes weaker, we observe the tendency that the regions show conductance signals comparable to or even higher than the six-fold points; the dot-reservoir coupling is empirically reduced by pinching the QPC gates and checking the conductance of the ordinary triple points [green points in Fig.~\ref{fig2}(d)]. The examples are shown in Fig.~\ref{fig4}(a) where the stripes with unusually high conductance connect the two neighboring six-fold points. The figure clearly shows that the conductance in the stripe is higher than the six-fold point and the triple points marked by arrows in the inset of Fig.~\ref{fig4}(a); note that the conductance of the triple points is above $\sim$ 0.25 $\times$ $e^{2}/h$ in Fig.~\ref{fig2}(b), while it is less than 0.1 $\times$ $e^{2}/h$ in Fig.~\ref{fig4}(a). Moreover, the conductance in the stripe is similar or even higher than that of the degeneracy points in Fig.~\ref{fig2}(b), although it is measured with relatively weaker electron tunneling between TQD and reservoirs than the case of Fig.~\ref{fig2}(b). We found that the conductance in the stripes is insensitive to temperature below 600 mK; see Fig.~\ref{fig4}(b). However, it is extremely sensitive to the P1 gate. The conductance in the stripe gets totally suppressed as $V_{P1}$ deviates from the value at which the P2-P3 plane shows the six-fold degeneracy points. At 52 mK, it vanishes totally when $V_{P1}$ deviates by 1 mV ($\sim$12 $\mu$eV in energy). This indicates that the stripes are strongly related to the charge frustration.

The features of the stripes may be partially understood by elastic cotunneling. In Fig.~\ref{fig4}(c), the energy levels of the TQD are calculated by using Eq.~(\ref{eq:eq1}) and by fitting to the experimental data. We find that the lowest excitations along the border between (0,0,1) and (1,0,0) [dashed red line in Fig.~\ref{fig2}(d)] are (0,0,0) and (1,0,1) states with excitation energy cost of about 25 $\mu$eV. In this situation, electrons can flow between reservoirs 1 and 3 through the TQD by cotunneling processes, for example, (i) such that the TQD is initially in the (0,0,1) state, (ii) that the TQD state is in the virtual state of (0,0,0) [or (1,0,1)] after an electron tunnels from QD3 to reservoir 3 [or from reservoir 1 to QD1], and (iii) finally that the TQD becomes (1,0,0) after an electron tunnels from reservoir 1 to QD1 [or from QD3 to reservoir 3]. Along the border, the energies of (0,0,1) and (1,0,0), the initial and the final states, are the same, hence, the cotunneling processes are elastic. As the elastic processes are insensitive to temperature \cite{PhysRevLett.65.2446, PhysRevLett.86.878}, they might explain the features in Fig.~\ref{fig4}(c). Moreover, they might explain as well the sensitivity to the change of $V_{P1}$. Our calculation shows that the change of $V_{P1}$ breaks the degeneracy between (0,0,1) and (1,0,0) along the border, and the resulting energy gap between (0,0,1) and (1,0,0) is comparable to the measurement temperature at which the stripes disappear. As the energy gap becomes larger than the temperature, the cotunneling processes become inelastic, hence, the conductance becomes suppressed, in good agreement with the experimental observation.
 However, it remains unclear why the conductance in the stripes is higher than that of the adjacent six-fold points. It is contrary to the naive expectation that the elastic cotunneling shows smaller conductance than the resonant transport of the six-fold degeneracy. One possible direction is to consider the combination effect of electron interactions and interference. There may be constructive or destructive interference between various trajectories due to the charge frustration around the six-fold points, which might enhance (suppress) conductance in the stripes (at the six-fold points). In addition, Kondo-type effects of isospins \cite{goldhaber1998kondo} might play a role. This part is left for future study.
 
In summary, the charge frustration appears in a triangular symmetric TQD. We reveal its nontrivial features in electron transport. This work provides a unique way of studying geometric frustration in a controllable way. It is also an important step towards spin frustration, quantum simulation, and quantum information processing in QDs.

\begin{acknowledgments}
We thank M. Heiblum for discussion and experimental support (access to his facilities). Also, we thank K. Kang and M.-S. Choi for discussions. This work was supported by the National Research Foundation of Korea (NRF) grants (No.2011-0003109, YC; 2011-0022955, HSS) and the Korea Science and Engineering Foundation (KOSEF) grant (No. 2010-0000268) funded by the Korea government (MEST).
\end{acknowledgments}


\begin{thebibliography}{24}%
\makeatletter
\providecommand \@ifxundefined [1]{%
 \@ifx{#1\undefined}
}%
\providecommand \@ifnum [1]{%
 \ifnum #1\expandafter \@firstoftwo
 \else \expandafter \@secondoftwo
 \fi
}%
\providecommand \@ifx [1]{%
 \ifx #1\expandafter \@firstoftwo
 \else \expandafter \@secondoftwo
 \fi
}%
\providecommand \natexlab [1]{#1}%
\providecommand \enquote  [1]{``#1''}%
\providecommand \bibnamefont  [1]{#1}%
\providecommand \bibfnamefont [1]{#1}%
\providecommand \citenamefont [1]{#1}%
\providecommand \href@noop [0]{\@secondoftwo}%
\providecommand \href [0]{\begingroup \@sanitize@url \@href}%
\providecommand \@href[1]{\@@startlink{#1}\@@href}%
\providecommand \@@href[1]{\endgroup#1\@@endlink}%
\providecommand \@sanitize@url [0]{\catcode `\\12\catcode `\$12\catcode
  `\&12\catcode `\#12\catcode `\^12\catcode `\_12\catcode `\%12\relax}%
\providecommand \@@startlink[1]{}%
\providecommand \@@endlink[0]{}%
\providecommand \url  [0]{\begingroup\@sanitize@url \@url }%
\providecommand \@url [1]{\endgroup\@href {#1}{\urlprefix }}%
\providecommand \urlprefix  [0]{URL }%
\providecommand \Eprint [0]{\href }%
\providecommand \doibase [0]{http://dx.doi.org/}%
\providecommand \selectlanguage [0]{\@gobble}%
\providecommand \bibinfo  [0]{\@secondoftwo}%
\providecommand \bibfield  [0]{\@secondoftwo}%
\providecommand \translation [1]{[#1]}%
\providecommand \BibitemOpen [0]{}%
\providecommand \bibitemStop [0]{}%
\providecommand \bibitemNoStop [0]{.\EOS\space}%
\providecommand \EOS [0]{\spacefactor3000\relax}%
\providecommand \BibitemShut  [1]{\csname bibitem#1\endcsname}%
\let\auto@bib@innerbib\@empty
\bibitem [{\citenamefont {Ong}\ and\ \citenamefont
  {Cava}(2004)}]{ong2004electronic}%
  \BibitemOpen
  \bibfield  {author} {\bibinfo {author} {\bibfnamefont {N.}~\bibnamefont
  {Ong}}\ and\ \bibinfo {author} {\bibfnamefont {R.}~\bibnamefont {Cava}},\
  }\href@noop {} {\bibfield  {journal} {\bibinfo  {journal} {Science}\ }\textbf
  {\bibinfo {volume} {305}},\ \bibinfo {pages} {52} (\bibinfo {year}
  {2004})}\BibitemShut {NoStop}%
\bibitem [{\citenamefont {Pauling}(1935)}]{pauling1935structure}%
  \BibitemOpen
  \bibfield  {author} {\bibinfo {author} {\bibfnamefont {L.}~\bibnamefont
  {Pauling}},\ }\href@noop {} {\bibfield  {journal} {\bibinfo  {journal} {J.
  Am. Chem. Soc.}\ }\textbf {\bibinfo {volume} {57}},\ \bibinfo {pages} {2680}
  (\bibinfo {year} {1935})}\BibitemShut {NoStop}%
\bibitem [{\citenamefont {Bramwell}\ and\ \citenamefont
  {Gingras}(2001)}]{bramwell2001spin}%
  \BibitemOpen
  \bibfield  {author} {\bibinfo {author} {\bibfnamefont {S.}~\bibnamefont
  {Bramwell}}\ and\ \bibinfo {author} {\bibfnamefont {M.}~\bibnamefont
  {Gingras}},\ }\href@noop {} {\bibfield  {journal} {\bibinfo  {journal}
  {Science}\ }\textbf {\bibinfo {volume} {294}},\ \bibinfo {pages} {1495}
  (\bibinfo {year} {2001})}\BibitemShut {NoStop}%
\bibitem [{\citenamefont {Ramirez}(1999)}]{ramirez1999superconductivity}%
  \BibitemOpen
  \bibfield  {author} {\bibinfo {author} {\bibfnamefont {A.}~\bibnamefont
  {Ramirez}},\ }\href@noop {} {\bibfield  {journal} {\bibinfo  {journal}
  {Nature}\ }\textbf {\bibinfo {volume} {399}},\ \bibinfo {pages} {527}
  (\bibinfo {year} {1999})}\BibitemShut {NoStop}%
\bibitem [{\citenamefont {Balents}(2010)}]{balents2010spin}%
  \BibitemOpen
  \bibfield  {author} {\bibinfo {author} {\bibfnamefont {L.}~\bibnamefont
  {Balents}},\ }\href@noop {} {\bibfield  {journal} {\bibinfo  {journal}
  {Nature}\ }\textbf {\bibinfo {volume} {464}},\ \bibinfo {pages} {199}
  (\bibinfo {year} {2010})}\BibitemShut {NoStop}%
\bibitem [{\citenamefont {Pratt}\ \emph {et~al.}(2011)\citenamefont {Pratt},
  \citenamefont {Baker}, \citenamefont {Blundell}, \citenamefont {Lancaster},
  \citenamefont {Ohira-Kawamura}, \citenamefont {Baines}, \citenamefont
  {Shimizu}, \citenamefont {Kanoda}, \citenamefont {Watanabe},\ and\
  \citenamefont {Saito}}]{pratt2011magnetic}%
  \BibitemOpen
  \bibfield  {author} {\bibinfo {author} {\bibfnamefont {F.}~\bibnamefont
  {Pratt}}, \bibinfo {author} {\bibfnamefont {P.}~\bibnamefont {Baker}},
  \bibinfo {author} {\bibfnamefont {S.}~\bibnamefont {Blundell}}, \bibinfo
  {author} {\bibfnamefont {T.}~\bibnamefont {Lancaster}}, \bibinfo {author}
  {\bibfnamefont {S.}~\bibnamefont {Ohira-Kawamura}}, \bibinfo {author}
  {\bibfnamefont {C.}~\bibnamefont {Baines}}, \bibinfo {author} {\bibfnamefont
  {Y.}~\bibnamefont {Shimizu}}, \bibinfo {author} {\bibfnamefont
  {K.}~\bibnamefont {Kanoda}}, \bibinfo {author} {\bibfnamefont
  {I.}~\bibnamefont {Watanabe}}, \ and\ \bibinfo {author} {\bibfnamefont
  {G.}~\bibnamefont {Saito}},\ }\href@noop {} {\bibfield  {journal} {\bibinfo
  {journal} {Nature}\ }\textbf {\bibinfo {volume} {471}},\ \bibinfo {pages}
  {612} (\bibinfo {year} {2011})}\BibitemShut {NoStop}%
\bibitem [{\citenamefont {Bramwell}\ \emph {et~al.}(2009)\citenamefont
  {Bramwell}, \citenamefont {Giblin}, \citenamefont {Calder}, \citenamefont
  {Aldus}, \citenamefont {Prabhakaran},\ and\ \citenamefont
  {Fennell}}]{bramwell2009measurement}%
  \BibitemOpen
  \bibfield  {author} {\bibinfo {author} {\bibfnamefont {S.}~\bibnamefont
  {Bramwell}}, \bibinfo {author} {\bibfnamefont {S.}~\bibnamefont {Giblin}},
  \bibinfo {author} {\bibfnamefont {S.}~\bibnamefont {Calder}}, \bibinfo
  {author} {\bibfnamefont {R.}~\bibnamefont {Aldus}}, \bibinfo {author}
  {\bibfnamefont {D.}~\bibnamefont {Prabhakaran}}, \ and\ \bibinfo {author}
  {\bibfnamefont {T.}~\bibnamefont {Fennell}},\ }\href@noop {} {\bibfield
  {journal} {\bibinfo  {journal} {Nature}\ }\textbf {\bibinfo {volume} {461}},\
  \bibinfo {pages} {956} (\bibinfo {year} {2009})}\BibitemShut {NoStop}%
\bibitem [{\citenamefont {Meirav}\ \emph {et~al.}(1990)\citenamefont {Meirav},
  \citenamefont {Kastner},\ and\ \citenamefont {Wind}}]{meirav1990single}%
  \BibitemOpen
  \bibfield  {author} {\bibinfo {author} {\bibfnamefont {U.}~\bibnamefont
  {Meirav}}, \bibinfo {author} {\bibfnamefont {M.}~\bibnamefont {Kastner}}, \
  and\ \bibinfo {author} {\bibfnamefont {S.}~\bibnamefont {Wind}},\ }\href@noop
  {} {\bibfield  {journal} {\bibinfo  {journal} {Phys. Rev. Lett.}\ }\textbf
  {\bibinfo {volume} {65}},\ \bibinfo {pages} {771} (\bibinfo {year}
  {1990})}\BibitemShut {NoStop}%
\bibitem [{\citenamefont {Tarucha}\ \emph {et~al.}(1996)\citenamefont
  {Tarucha}, \citenamefont {Austing}, \citenamefont {Honda}, \citenamefont
  {van~der Hage},\ and\ \citenamefont {Kouwenhoven}}]{PhysRevLett.77.3613}%
  \BibitemOpen
  \bibfield  {author} {\bibinfo {author} {\bibfnamefont {S.}~\bibnamefont
  {Tarucha}}, \bibinfo {author} {\bibfnamefont {D.~G.}\ \bibnamefont
  {Austing}}, \bibinfo {author} {\bibfnamefont {T.}~\bibnamefont {Honda}},
  \bibinfo {author} {\bibfnamefont {R.~J.}\ \bibnamefont {van~der Hage}}, \
  and\ \bibinfo {author} {\bibfnamefont {L.~P.}\ \bibnamefont {Kouwenhoven}},\
  }\href {\doibase 10.1103/PhysRevLett.77.3613} {\bibfield  {journal} {\bibinfo
   {journal} {Phys. Rev. Lett.}\ }\textbf {\bibinfo {volume} {77}},\ \bibinfo
  {pages} {3613} (\bibinfo {year} {1996})}\BibitemShut {NoStop}%
\bibitem [{\citenamefont {Livermore}\ \emph {et~al.}(1996)\citenamefont
  {Livermore}, \citenamefont {Crouch}, \citenamefont {Westervelt},
  \citenamefont {Campman}, \citenamefont {Gossard} \emph
  {et~al.}}]{livermore1996coulomb}%
  \BibitemOpen
  \bibfield  {author} {\bibinfo {author} {\bibfnamefont {C.}~\bibnamefont
  {Livermore}}, \bibinfo {author} {\bibfnamefont {C.}~\bibnamefont {Crouch}},
  \bibinfo {author} {\bibfnamefont {R.}~\bibnamefont {Westervelt}}, \bibinfo
  {author} {\bibfnamefont {K.}~\bibnamefont {Campman}}, \bibinfo {author}
  {\bibfnamefont {A.}~\bibnamefont {Gossard}},  \emph {et~al.},\ }\href@noop {}
  {\bibfield  {journal} {\bibinfo  {journal} {Science}\ }\textbf {\bibinfo
  {volume} {274}},\ \bibinfo {pages} {1332} (\bibinfo {year}
  {1996})}\BibitemShut {NoStop}%
\bibitem [{\citenamefont {Goldhaber-Gordon}\ \emph {et~al.}(1998)\citenamefont
  {Goldhaber-Gordon}, \citenamefont {Shtrikman}, \citenamefont {Mahalu},
  \citenamefont {Abusch-Magder}, \citenamefont {Meirav},\ and\ \citenamefont
  {Kastner}}]{goldhaber1998kondo}%
  \BibitemOpen
  \bibfield  {author} {\bibinfo {author} {\bibfnamefont {D.}~\bibnamefont
  {Goldhaber-Gordon}}, \bibinfo {author} {\bibfnamefont {H.}~\bibnamefont
  {Shtrikman}}, \bibinfo {author} {\bibfnamefont {D.}~\bibnamefont {Mahalu}},
  \bibinfo {author} {\bibfnamefont {D.}~\bibnamefont {Abusch-Magder}}, \bibinfo
  {author} {\bibfnamefont {U.}~\bibnamefont {Meirav}}, \ and\ \bibinfo {author}
  {\bibfnamefont {M.}~\bibnamefont {Kastner}},\ }\href@noop {} {\bibfield
  {journal} {\bibinfo  {journal} {Nature}\ }\textbf {\bibinfo {volume} {391}},\
  \bibinfo {pages} {156} (\bibinfo {year} {1998})}\BibitemShut {NoStop}%
\bibitem [{\citenamefont {Van~der Wiel}\ \emph {et~al.}(2000)\citenamefont
  {Van~der Wiel}, \citenamefont {De~Franceschi}, \citenamefont {Fujisawa},
  \citenamefont {Elzerman}, \citenamefont {Tarucha},\ and\ \citenamefont
  {Kouwenhoven}}]{van2000kondo}%
  \BibitemOpen
  \bibfield  {author} {\bibinfo {author} {\bibfnamefont {W.}~\bibnamefont
  {Van~der Wiel}}, \bibinfo {author} {\bibfnamefont {S.}~\bibnamefont
  {De~Franceschi}}, \bibinfo {author} {\bibfnamefont {T.}~\bibnamefont
  {Fujisawa}}, \bibinfo {author} {\bibfnamefont {J.}~\bibnamefont {Elzerman}},
  \bibinfo {author} {\bibfnamefont {S.}~\bibnamefont {Tarucha}}, \ and\
  \bibinfo {author} {\bibfnamefont {L.}~\bibnamefont {Kouwenhoven}},\
  }\href@noop {} {\bibfield  {journal} {\bibinfo  {journal} {Science}\ }\textbf
  {\bibinfo {volume} {289}},\ \bibinfo {pages} {2105} (\bibinfo {year}
  {2000})}\BibitemShut {NoStop}%
\bibitem [{\citenamefont {Koppens}\ \emph {et~al.}(2006)\citenamefont
  {Koppens}, \citenamefont {Buizert}, \citenamefont {Tielrooij}, \citenamefont
  {Vink}, \citenamefont {Nowack}, \citenamefont {Meunier}, \citenamefont
  {Kouwenhoven},\ and\ \citenamefont {Vandersypen}}]{koppens2006driven}%
  \BibitemOpen
  \bibfield  {author} {\bibinfo {author} {\bibfnamefont {F.}~\bibnamefont
  {Koppens}}, \bibinfo {author} {\bibfnamefont {C.}~\bibnamefont {Buizert}},
  \bibinfo {author} {\bibfnamefont {K.}~\bibnamefont {Tielrooij}}, \bibinfo
  {author} {\bibfnamefont {I.}~\bibnamefont {Vink}}, \bibinfo {author}
  {\bibfnamefont {K.}~\bibnamefont {Nowack}}, \bibinfo {author} {\bibfnamefont
  {T.}~\bibnamefont {Meunier}}, \bibinfo {author} {\bibfnamefont
  {L.}~\bibnamefont {Kouwenhoven}}, \ and\ \bibinfo {author} {\bibfnamefont
  {L.}~\bibnamefont {Vandersypen}},\ }\href@noop {} {\bibfield  {journal}
  {\bibinfo  {journal} {Nature}\ }\textbf {\bibinfo {volume} {442}},\ \bibinfo
  {pages} {766} (\bibinfo {year} {2006})}\BibitemShut {NoStop}%
\bibitem [{\citenamefont {Brunner}\ \emph {et~al.}(2011)\citenamefont
  {Brunner}, \citenamefont {Shin}, \citenamefont {Obata}, \citenamefont
  {Pioro-Ladri\`ere}, \citenamefont {Kubo}, \citenamefont {Yoshida},
  \citenamefont {Taniyama}, \citenamefont {Tokura},\ and\ \citenamefont
  {Tarucha}}]{PhysRevLett.107.146801}%
  \BibitemOpen
  \bibfield  {author} {\bibinfo {author} {\bibfnamefont {R.}~\bibnamefont
  {Brunner}}, \bibinfo {author} {\bibfnamefont {Y.-S.}\ \bibnamefont {Shin}},
  \bibinfo {author} {\bibfnamefont {T.}~\bibnamefont {Obata}}, \bibinfo
  {author} {\bibfnamefont {M.}~\bibnamefont {Pioro-Ladri\`ere}}, \bibinfo
  {author} {\bibfnamefont {T.}~\bibnamefont {Kubo}}, \bibinfo {author}
  {\bibfnamefont {K.}~\bibnamefont {Yoshida}}, \bibinfo {author} {\bibfnamefont
  {T.}~\bibnamefont {Taniyama}}, \bibinfo {author} {\bibfnamefont
  {Y.}~\bibnamefont {Tokura}}, \ and\ \bibinfo {author} {\bibfnamefont
  {S.}~\bibnamefont {Tarucha}},\ }\href {\doibase
  10.1103/PhysRevLett.107.146801} {\bibfield  {journal} {\bibinfo  {journal}
  {Phys. Rev. Lett.}\ }\textbf {\bibinfo {volume} {107}},\ \bibinfo {pages}
  {146801} (\bibinfo {year} {2011})}\BibitemShut {NoStop}%
\bibitem [{\citenamefont {Shulman}\ \emph {et~al.}(2012)\citenamefont
  {Shulman}, \citenamefont {Dial}, \citenamefont {Harvey}, \citenamefont
  {Bluhm}, \citenamefont {Umansky},\ and\ \citenamefont
  {Yacoby}}]{shulman2012demonstration}%
  \BibitemOpen
  \bibfield  {author} {\bibinfo {author} {\bibfnamefont {M.}~\bibnamefont
  {Shulman}}, \bibinfo {author} {\bibfnamefont {O.}~\bibnamefont {Dial}},
  \bibinfo {author} {\bibfnamefont {S.}~\bibnamefont {Harvey}}, \bibinfo
  {author} {\bibfnamefont {H.}~\bibnamefont {Bluhm}}, \bibinfo {author}
  {\bibfnamefont {V.}~\bibnamefont {Umansky}}, \ and\ \bibinfo {author}
  {\bibfnamefont {A.}~\bibnamefont {Yacoby}},\ }\href@noop {} {\bibfield
  {journal} {\bibinfo  {journal} {Science}\ }\textbf {\bibinfo {volume}
  {336}},\ \bibinfo {pages} {202} (\bibinfo {year} {2012})}\BibitemShut
  {NoStop}%
\bibitem [{\citenamefont {Vidan}\ \emph {et~al.}(2004)\citenamefont {Vidan},
  \citenamefont {Westervelt}, \citenamefont {Stopa}, \citenamefont {Hanson},\
  and\ \citenamefont {Gossard}}]{vidan2004triple}%
  \BibitemOpen
  \bibfield  {author} {\bibinfo {author} {\bibfnamefont {A.}~\bibnamefont
  {Vidan}}, \bibinfo {author} {\bibfnamefont {R.}~\bibnamefont {Westervelt}},
  \bibinfo {author} {\bibfnamefont {M.}~\bibnamefont {Stopa}}, \bibinfo
  {author} {\bibfnamefont {M.}~\bibnamefont {Hanson}}, \ and\ \bibinfo {author}
  {\bibfnamefont {A.}~\bibnamefont {Gossard}},\ }\href@noop {} {\bibfield
  {journal} {\bibinfo  {journal} {Applied physics letters}\ }\textbf {\bibinfo
  {volume} {85}},\ \bibinfo {pages} {3602} (\bibinfo {year}
  {2004})}\BibitemShut {NoStop}%
\bibitem [{\citenamefont {Schr{\"o}er}\ \emph {et~al.}(2007)\citenamefont
  {Schr{\"o}er}, \citenamefont {Greentree}, \citenamefont {Gaudreau},
  \citenamefont {Eberl}, \citenamefont {Hollenberg}, \citenamefont {Kotthaus},\
  and\ \citenamefont {Ludwig}}]{schroer2007electrostatically}%
  \BibitemOpen
  \bibfield  {author} {\bibinfo {author} {\bibfnamefont {D.}~\bibnamefont
  {Schr{\"o}er}}, \bibinfo {author} {\bibfnamefont {A.}~\bibnamefont
  {Greentree}}, \bibinfo {author} {\bibfnamefont {L.}~\bibnamefont {Gaudreau}},
  \bibinfo {author} {\bibfnamefont {K.}~\bibnamefont {Eberl}}, \bibinfo
  {author} {\bibfnamefont {L.}~\bibnamefont {Hollenberg}}, \bibinfo {author}
  {\bibfnamefont {J.}~\bibnamefont {Kotthaus}}, \ and\ \bibinfo {author}
  {\bibfnamefont {S.}~\bibnamefont {Ludwig}},\ }\href@noop {} {\bibfield
  {journal} {\bibinfo  {journal} {Phys. Rev. B}\ }\textbf {\bibinfo {volume}
  {76}},\ \bibinfo {pages} {075306} (\bibinfo {year} {2007})}\BibitemShut
  {NoStop}%
\bibitem [{\citenamefont {Delgado}\ \emph {et~al.}(2008)\citenamefont
  {Delgado}, \citenamefont {Shim}, \citenamefont {Korkusinski}, \citenamefont
  {Gaudreau}, \citenamefont {Studenikin}, \citenamefont {Sachrajda},\ and\
  \citenamefont {Hawrylak}}]{delgado2008spin}%
  \BibitemOpen
  \bibfield  {author} {\bibinfo {author} {\bibfnamefont {F.}~\bibnamefont
  {Delgado}}, \bibinfo {author} {\bibfnamefont {Y.}~\bibnamefont {Shim}},
  \bibinfo {author} {\bibfnamefont {M.}~\bibnamefont {Korkusinski}}, \bibinfo
  {author} {\bibfnamefont {L.}~\bibnamefont {Gaudreau}}, \bibinfo {author}
  {\bibfnamefont {S.}~\bibnamefont {Studenikin}}, \bibinfo {author}
  {\bibfnamefont {A.}~\bibnamefont {Sachrajda}}, \ and\ \bibinfo {author}
  {\bibfnamefont {P.}~\bibnamefont {Hawrylak}},\ }\href@noop {} {\bibfield
  {journal} {\bibinfo  {journal} {Phys. Rev. Lett.}\ }\textbf {\bibinfo
  {volume} {101}},\ \bibinfo {pages} {226810} (\bibinfo {year}
  {2008})}\BibitemShut {NoStop}%
\bibitem [{\citenamefont {Nowack}\ \emph {et~al.}(2007)\citenamefont {Nowack},
  \citenamefont {Koppens}, \citenamefont {Nazarov},\ and\ \citenamefont
  {Vandersypen}}]{nowack2007coherent}%
  \BibitemOpen
  \bibfield  {author} {\bibinfo {author} {\bibfnamefont {K.}~\bibnamefont
  {Nowack}}, \bibinfo {author} {\bibfnamefont {F.}~\bibnamefont {Koppens}},
  \bibinfo {author} {\bibfnamefont {Y.}~\bibnamefont {Nazarov}}, \ and\
  \bibinfo {author} {\bibfnamefont {L.}~\bibnamefont {Vandersypen}},\
  }\href@noop {} {\bibfield  {journal} {\bibinfo  {journal} {Science}\ }\textbf
  {\bibinfo {volume} {318}},\ \bibinfo {pages} {1430} (\bibinfo {year}
  {2007})}\BibitemShut {NoStop}%
\bibitem [{\citenamefont {Granger}\ \emph {et~al.}(2010)\citenamefont
  {Granger}, \citenamefont {Gaudreau}, \citenamefont {Kam}, \citenamefont
  {Pioro-Ladri\`ere}, \citenamefont {Studenikin}, \citenamefont {Wasilewski},
  \citenamefont {Zawadzki},\ and\ \citenamefont
  {Sachrajda}}]{PhysRevB.82.075304}%
  \BibitemOpen
  \bibfield  {author} {\bibinfo {author} {\bibfnamefont {G.}~\bibnamefont
  {Granger}}, \bibinfo {author} {\bibfnamefont {L.}~\bibnamefont {Gaudreau}},
  \bibinfo {author} {\bibfnamefont {A.}~\bibnamefont {Kam}}, \bibinfo {author}
  {\bibfnamefont {M.}~\bibnamefont {Pioro-Ladri\`ere}}, \bibinfo {author}
  {\bibfnamefont {S.~A.}\ \bibnamefont {Studenikin}}, \bibinfo {author}
  {\bibfnamefont {Z.~R.}\ \bibnamefont {Wasilewski}}, \bibinfo {author}
  {\bibfnamefont {P.}~\bibnamefont {Zawadzki}}, \ and\ \bibinfo {author}
  {\bibfnamefont {A.~S.}\ \bibnamefont {Sachrajda}},\ }\href {\doibase
  10.1103/PhysRevB.82.075304} {\bibfield  {journal} {\bibinfo  {journal} {Phys.
  Rev. B}\ }\textbf {\bibinfo {volume} {82}},\ \bibinfo {pages} {075304}
  (\bibinfo {year} {2010})}\BibitemShut {NoStop}%
\bibitem [{\citenamefont {Gaudreau}\ \emph {et~al.}(2011)\citenamefont
  {Gaudreau}, \citenamefont {Granger}, \citenamefont {Kam}, \citenamefont
  {Aers}, \citenamefont {Studenikin}, \citenamefont {Zawadzki}, \citenamefont
  {Pioro-Ladri{\`e}re}, \citenamefont {Wasilewski},\ and\ \citenamefont
  {Sachrajda}}]{gaudreau2011coherent}%
  \BibitemOpen
  \bibfield  {author} {\bibinfo {author} {\bibfnamefont {L.}~\bibnamefont
  {Gaudreau}}, \bibinfo {author} {\bibfnamefont {G.}~\bibnamefont {Granger}},
  \bibinfo {author} {\bibfnamefont {A.}~\bibnamefont {Kam}}, \bibinfo {author}
  {\bibfnamefont {G.}~\bibnamefont {Aers}}, \bibinfo {author} {\bibfnamefont
  {S.}~\bibnamefont {Studenikin}}, \bibinfo {author} {\bibfnamefont
  {P.}~\bibnamefont {Zawadzki}}, \bibinfo {author} {\bibfnamefont
  {M.}~\bibnamefont {Pioro-Ladri{\`e}re}}, \bibinfo {author} {\bibfnamefont
  {Z.}~\bibnamefont {Wasilewski}}, \ and\ \bibinfo {author} {\bibfnamefont
  {A.}~\bibnamefont {Sachrajda}},\ }\href@noop {} {\bibfield  {journal}
  {\bibinfo  {journal} {Nature Physics}\ } (\bibinfo {year}
  {2011})}\BibitemShut {NoStop}%
\bibitem [{\citenamefont {Kretinin}\ and\ \citenamefont
  {Chung}(2012)}]{kretinin2012wide}%
  \BibitemOpen
  \bibfield  {author} {\bibinfo {author} {\bibfnamefont {A.}~\bibnamefont
  {Kretinin}}\ and\ \bibinfo {author} {\bibfnamefont {Y.}~\bibnamefont
  {Chung}},\ }\href@noop {} {\bibfield  {journal} {\bibinfo  {journal} {Rev.
  Sci. Instrum.}\ }\textbf {\bibinfo {volume} {83}},\ \bibinfo {pages} {084704}
  (\bibinfo {year} {2012})}\BibitemShut {NoStop}%
\bibitem [{\citenamefont {Averin}\ and\ \citenamefont
  {Nazarov}(1990)}]{PhysRevLett.65.2446}%
  \BibitemOpen
  \bibfield  {author} {\bibinfo {author} {\bibfnamefont {D.~V.}\ \bibnamefont
  {Averin}}\ and\ \bibinfo {author} {\bibfnamefont {Y.~V.}\ \bibnamefont
  {Nazarov}},\ }\href {\doibase 10.1103/PhysRevLett.65.2446} {\bibfield
  {journal} {\bibinfo  {journal} {Phys. Rev. Lett.}\ }\textbf {\bibinfo
  {volume} {65}},\ \bibinfo {pages} {2446} (\bibinfo {year}
  {1990})}\BibitemShut {NoStop}%
\bibitem [{\citenamefont {De~Franceschi}\ \emph {et~al.}(2001)\citenamefont
  {De~Franceschi}, \citenamefont {Sasaki}, \citenamefont {Elzerman},
  \citenamefont {van~der Wiel}, \citenamefont {Tarucha},\ and\ \citenamefont
  {Kouwenhoven}}]{PhysRevLett.86.878}%
  \BibitemOpen
  \bibfield  {author} {\bibinfo {author} {\bibfnamefont {S.}~\bibnamefont
  {De~Franceschi}}, \bibinfo {author} {\bibfnamefont {S.}~\bibnamefont
  {Sasaki}}, \bibinfo {author} {\bibfnamefont {J.~M.}\ \bibnamefont
  {Elzerman}}, \bibinfo {author} {\bibfnamefont {W.~G.}\ \bibnamefont {van~der
  Wiel}}, \bibinfo {author} {\bibfnamefont {S.}~\bibnamefont {Tarucha}}, \ and\
  \bibinfo {author} {\bibfnamefont {L.~P.}\ \bibnamefont {Kouwenhoven}},\
  }\href {\doibase 10.1103/PhysRevLett.86.878} {\bibfield  {journal} {\bibinfo
  {journal} {Phys. Rev. Lett.}\ }\textbf {\bibinfo {volume} {86}},\ \bibinfo
  {pages} {878} (\bibinfo {year} {2001})}\BibitemShut {NoStop}%
\end{thebibliography}
%
\end{document}